\documentclass[letter]{aa}  

\usepackage{graphicx}
\usepackage{amsmath}
\usepackage{stmaryrd}
\usepackage[dvipsnames]{pstricks}
\usepackage{txfonts}
\usepackage{xcolor}
\usepackage{booktabs}
\usepackage{natbib}
\bibpunct{(}{)}{;}{a}{}{,} 
\usepackage[colorlinks=True, 
            urlcolor=blue,
            linkcolor=blue,
            citecolor= blue,
            dvips]{hyperref}
\usepackage{wasysym}

\newcommand{\disp}[1]{\displaystyle #1}

\newcommand{\Amp}{{\cal A}}
\newcommand{\Fbot}{F_{\hbox{\scriptsize bot}}}

\newcommand{\Tbump}{T_{\hbox{\scriptsize bump}}}
\newcommand{\Kmin}{K_{\hbox{\scriptsize min}}}
\newcommand{\Kmax}{K_{\hbox{\scriptsize max}}}
\newcommand{\rhotop}{\rho_{\hbox{\scriptsize top}}}
\newcommand{\Ttop}{T_{\hbox{\scriptsize top}}}

\renewcommand{\na}{ \vec{\nabla} }

\def\Div{\mathop{\hbox{div}}\nolimits}
\newcommand{\lp}{ \left(}
\newcommand{\rp}{ \right)}

\newcommand{\Et}{{\cal E}_{t}}
\newcommand{\Ft}{{\cal F}_{\text{kin}}}
\newcommand{\Fc}{{\cal F}_{\text{conv}}}
\newcommand{\Fr}{{\cal F}_{\text{rad}}}
\newcommand{\Fs}{{\cal F}_{\text{mod}}}

\newcommand{\Fst}{{\cal F}_{\text{St}}}
\newcommand{\Fku}{{\cal F}_{\text{Ku}}}
\newcommand{\alpst}{\alpha_{\text{St}}}
\newcommand{\alpku}{\alpha_{\text{Ku}}}
\newcommand{\gradad}{\nabla_{\text{ad}}}
\DeclareMathOperator{\sign}{sign}

\begin{document}

\title{A test of time-dependent theories of stellar convection}

\author{T.\ Gastine\inst{1} \and B.\ Dintrans\inst{2}}
\institute{Max-Planck-Institut f\"ur Sonnensystemforschung, Max-Planck-Strasse 
2, 37191 Katlenburg-Lindau, Germany  \email{gastine@mps.mpg.de}\and Institut de Recherche en Astrophysique 
et Plan\'etologie, CNRS/Universit\'e de Toulouse, 14 Av. Edouard Belin, 31400 
Toulouse, France} 

\date{March 16, 2011,~ accepted for publication}

\abstract
{In Cepheids close to the red edge of the classical instability strip, a
coupling occurs between the acoustic oscillations and the convective
motions close to the surface.
The best topical models that account for this coupling rely on 1-D
time-dependent convection (TDC) formulations. However, their intrinsic
weakness comes from the
large number of unconstrained free parameters entering in the description of
turbulent convection.}
{We compare two widely used TDC models with the first
two-dimensional nonlinear direct numerical simulations (DNS) of the
convection-pulsation coupling in which the acoustic oscillations are 
self-sustained by the $\kappa$-mechanism.}
{The free parameters appearing in the Stellingwerf and Kuhfu\ss~TDC
recipes are constrained using a $\chi^2$-test with the time-dependent 
convective flux that evolves in nonlinear simulations of highly-compressible 
convection with $\kappa$-mechanism.}
{This work emphasises some inherent limits of TDC
models, that is, the temporal variability and non-universality of their
free parameters. More importantly, within these limits, Stellingwerf's
formalism is found to give better spatial and temporal agreements with 
the nonlinear simulation than Kuhfu\ss's one.
It may therefore be
preferred in 1-D TDC hydrocodes or stellar evolution codes.}
{}


\maketitle

\section{Introduction}

The first theoretical calculations of the Cepheids instability strip
done in the 60's assumed that convection was steady with respect to 
oscillations. Unfortunately, this ``frozen-in
convection'' approximation led to a cooler red edge than the observed
one as the strong coupling between convection and pulsations occurring in cool 
Cepheids was ignored \citep[e.g.][]{Baker65}. Then, following the pioneering 
works of
\cite{Unno} and \cite{Gough}, several time-dependent convection (TDC) models
have been developed to investigate the influence of the
convection onto the pulsational stability \citep[e.g.][]{St,K, GW}.
The last up-to-date TDC models actually succeed
in predicting both a red edge close to the observed one and
realistic luminosity curves \citep[e.g.][]{YKB98, Bono, Feuch99,
Koll02}.

However, all of these TDC models suffer from a common weakness due to
the numerous free parameters, usually known as $\alpha$ coefficients, that
describe the turbulent convection
\citep[e.g. the 8 dimensionless parameters in][]{Smolec2010}.
These parameters are either obtained from a fit to the observations or
hardly constrained by theory when taking the asymptotic limit
of stationary convection \citep{BV1, BV2}. A parametric study carried
out by \cite{YKB98} has emphasised the intrinsic degeneracy of TDC
models as similar instability strips have been obtained with different
sets of parameters.

Direct numerical simulations (hereafter DNS) are able to constrain these
TDC models as they fully account for the nonlinearities involved
in the convection-pulsation coupling. These nonlinear simulations
are challenging as they require both
large-amplitude oscillations and convective motions.
In our last 2-D simulations, we have improved the 
way acoustic waves are generated by reproducing the self-consistent
excitation operating in variable stars, that is, the $\kappa$-mechanism
\citep[][hereafter GD2011]{paperIII}. The resulting coupling of acoustic modes
with convection is therefore more consistent as the mode amplitude is not
imposed artificially. We have shown in GD2011 that the convective plumes may
either quench the radial oscillations or coexist with the acoustic modes,
depending mainly on the density contrast of the equilibrium model.

The purpose of this letter is to compare the fully nonlinear results
with two main TDC models: (\textit{i}) the first one refers to an
initial formulation of \cite{St} that has been used and improved by
\cite{BS} and \cite{Bono}; (\textit{ii}) the other one has been developed
by \cite{K} and \cite{GW} and is implemented both in the Vienna hydrocode
\citep[e.g.][]{WF,Feuch99} and in the Florida-Budapest one
\citep[e.g][]{YKB98,Koll02}. In these models, a single equation for the
turbulent kinetic energy $\Et$ is added to the classical mean-field equations
and the main second-order correlations, as for example the convective flux,
are expressed as a function of $\Et$ only \citep[see e.g.][]{Baker, GW2,
Buchler-ASP, Buchler09}.

We investigate here in more details a particular simulation where the
oscillations strongly modulate the convective flux. The nonlinear results are
first compared at each snapshot with the TDC recipes by computing a
$\chi^2$-statistics of the relevant $\alpha$ coefficients. Secondly, the mean values of
$\alpha$ are used to compare the optimal TDC fluxes with the DNS ones.
The formulation of Stellingwerf is found to be closer to the nonlinear
result than Kuhfu\ss's one:  (\textit{i}) the temporal variation of the
$\alpha$ coefficient is weaker; (\textit{ii}) the mean convective
flux is closer to the DNS one, especially in the description of the
overshooting area.

\section{The hydrodynamic model}

We consider a local 2-D box of size $L_x\times L_z$, filled by a
perfect monatomic gas, and centered on both sides of an ionisation
region, of which the associated opacity bump is shaped by a hollow in the
temperature-dependent radiative conductivity profile $K(T)$. Such a
configuration can lead to unstable acoustic modes due to the driving
term $\na \cdot K(T)\na T$ in the energy equation, i.e. this is
the $\kappa$-mechanism \citep{paperI, paperII}. Furthermore, this hollow
in $K(T)$ is deep enough such as the equilibrium temperature gradient is
locally superadiabatic and convective motions develop here according to
Schwarzschild's criterion.

The governing hydrodynamic equations are written in nondimensional
form by choosing $L_z$ as the length scale and $\sqrt{c_p\Ttop}$
as the velocity one, hence the time scale $L_z/\sqrt{c_p\Ttop}$ (with $c_p$
the specific heat and $\Ttop$ the surface temperature).
The resulting nonlinear set of equations is advanced in time
with the high-order finite-difference pencil code\footnote{See 
\url{http://www.nordita.org/software/pencil-code/} and
\cite{Pencil-Code}.}, which is fully explicit except for the radiative
diffusion term that is solved implicitly thanks to a parallel alternate
direction implicit (ADI) solver. With the chosen units, the simulation box
spans about $10\%$ of the star radius on both sides of the ionisation region while
the timestep is about 1 minute, such that a simulation typically spans over 
4500 days (see GD2011).

\section{Results}

The time evolution of the convective flux obtained in fully nonlinear
2-D simulations is compared with the following TDC expressions developed
by \cite{St} and \cite{K}:

\begin{equation}
\left\lbrace
\begin{aligned}
 \Fst(z,t) & = \alpst \dfrac{A}{B} \Et \sign(\nabla
-\gradad)\sqrt{|\nabla -\gradad |}, \\ 
 \Fku(z,t) & = \alpku A \sqrt{\Et}\lp \nabla -\gradad \rp,
\end{aligned}
 \right.
 \label{eq:MLT}
\end{equation}
where $\nabla=d\ln T/d\ln p$, $\gradad=1-c_v/c_p$ and

\begin{equation}
\Et(z,t) =\left\langle\dfrac{{u'_z}^2}{2}\right \rangle,\
A = c_p\left\langle \rho \right\rangle \left\langle T \right\rangle
\text{ and } B= \sqrt{c_p\left\langle T \right\rangle \gradad},
\label{eq:eturb}
\end{equation}
with $p$ and $\rho$ the pressure and density, respectively, and
the brackets denote an horizontal average. Two free dimensionless
parameters $\alpst$ and $\alpku$ are introduced that the simulations allow
to constrain. We first focus on a 2-D DNS that is similar
to the G8 simulation in GD2011, that is, a simulation in which the total
kinetic energy is almost entirely contained in acoustic modes excited by
$\kappa$-mechanism (80\%), the rest being in convective plumes (20\%).
The convective motions do not affect much the growth and the nonlinear
saturation of the unstable radial acoustic mode. We can therefore expect
that the heat transport is strongly modulated by wave motions, what is an
ideal frame for testing the accuracy of time-dependent convection models.

\subsection{Temporal modulation of the convective flux in the DNS}

\begin{figure}
 \centering
 \includegraphics[width=9cm]{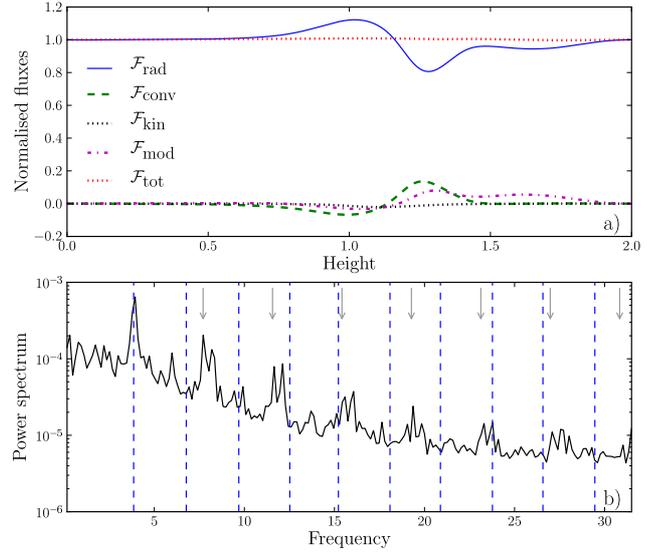}
 \caption{\textbf{a)} Mean vertical profiles of radiative $\Fr$
(solid blue line), turbulent enthalpy $\Fc$ (dashed green line), kinetic
$\Ft$ (dotted black line), modes $\Fs$ (dot-dashed magenta line) and total
${\cal F}_{\text{tot}}$ (dotted red line) fluxes, normalised to the bottom 
flux $\Fbot$. \textbf{b)} Temporal power spectrum for the convective flux 
only.}
 \label{fig:fluxes}
\end{figure}

The imposed bottom flux $\Fbot$ is mainly transported through the
computational domain by the radiative $\Fr$, enthalpy and kinetic $\Ft$
fluxes. Following GD2011, the enthalpy flux is divided into the classical
convective flux $\Fc$ and the contribution coming from the (unstable)
fundamental mode (hereafter $\Fs$) with

\begin{equation}
\left\lbrace
\begin{array}{lll}
\Fc(z,t) & = c_p \left\langle \rho u_z T' \right\rangle,\quad
& \Fs(z,t) = c_p \left\langle \rho u_z\right\rangle\theta, \\ \\
\Fr(z,t) & = -\left\langle K(T) \nabla T \right\rangle,\quad
& \Ft(z,t) = \dfrac{1}{2}\left\langle \rho u_z u^2 \right\rangle,
\end{array}\right.
\label{eq:fluxes}
\end{equation}
where the primed quantities denote the fluctuations about the horizontal
average and $\theta$ is the temperature eigenfunction of the fundamental
mode. The resulting time-averaged and normalised fluxes are given in
Fig.~\ref{fig:fluxes}a.

The bulk of the total flux is transported by the
radiative flux, except in the convective zone where $\Fc/\Fbot\simeq
20\%$, while the kinetic flux is negligible ($\Ft/\Fbot\leq
1\%$). Concerning $\Fs$, one notes that it is hardly as large as $\Fc$
in the convective zone. This quantity is a
good signature of the amplitude of the acoustic modes as the higher $\Fs$,
the larger the radial oscillations (GD2011), therefore confirming the efficiency
of the $\kappa$-mechanism in this simulation.

The signature of the temporal modulation of the convective flux
is extracted from its Fourier spectrum in time, that is, we first
compute $\widehat{\cal F}_{\text{conv}}(z,\omega)$, with $\omega$
the frequency, and second integrate over the vertical direction $z$
to get the power spectrum $\widehat{\cal F}_{\text{conv}}(\omega)$
(Fig.~\ref{fig:fluxes}b). Several discrete peaks appear about given
frequencies, of which the physical nature is emphasised after superimposing
the linear acoustic eigenfrequencies (the vertical dashed blue lines). The
matching with the fundamental mode frequency $\omega_{00}=3.85$ is
perfect while the weak-amplitude secondary peaks rather correspond to
harmonics of $\omega_{00}$ (i.e. $2\omega_{00},3\omega_{00},\dots$,
shaped by downward-directed vertical gray arrows in
Fig.~\ref{fig:fluxes}b) than to the acoustic overtones $\omega_{01},\
\omega_{02},\dots$. It means that the amplitude of the fundamental
acoustic mode is large enough to generate several harmonics through a
nonlinear cascade and this is again an indication of both the
robustness of the $\kappa$-driving and the relevance of this kind of
convection-pulsation simulation to check the TDC recipes.

\begin{figure*}
 \centering
 \includegraphics[width=18cm]{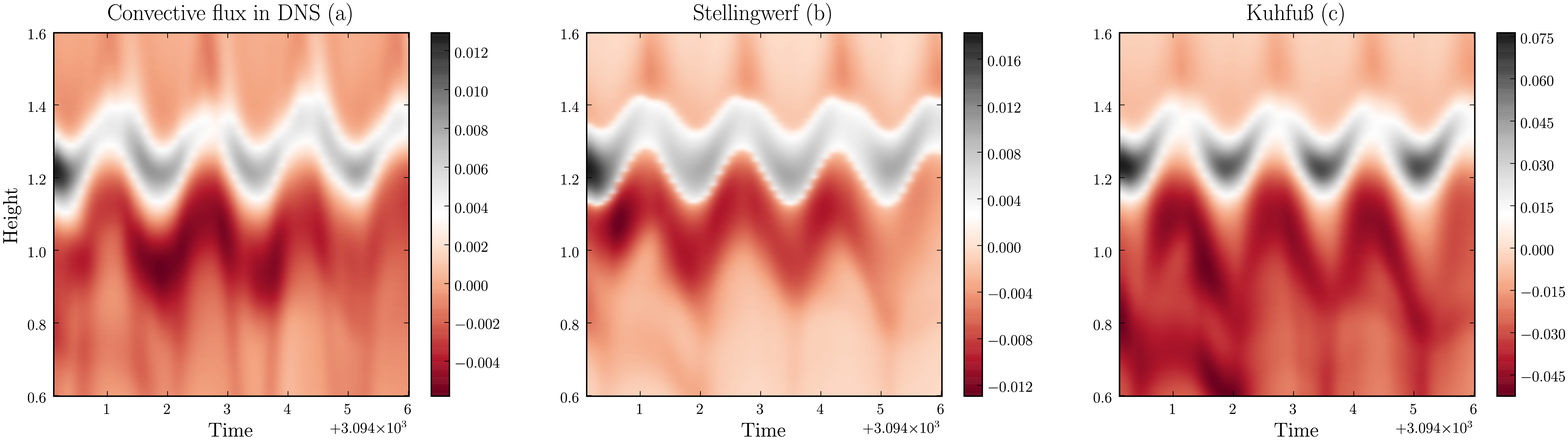}
 \caption{Evolution of the convective flux in a $(t,\ z)$ plane: (\textbf{a}) in the DNS; (\textbf{b}) with Stellingwerf's formalism $\Fst$;
(\textbf{c}) with Kuhfu\ss's~formalism $\Fku$ (for 
$\alpst=\alpku=1$). The 
vertical extent is centered on both sides of the convection zone to emphasise its oscillations.
These snapshots are computed after 1800 periods of oscillations, i.e.  well 
after the
nonlinear saturation is achieved.}
 \label{fig:contours}
\end{figure*}

\subsection{DNS vs TDC models: convective patterns}

We first compare the time evolution of the horizontally averaged
convective flux obtained in the DNS (Eq.~\ref{eq:fluxes}) with its
TDC counterparts $\Fst$ and $\Fku$ (Eq.~\ref{eq:MLT}). As we are interested here in the
qualitative agreement between the convective patterns in the
plane $(t,\ z)$, we simply assume $\alpst=\alpku =1$.

The three resulting patterns are displayed in Fig.~\ref{fig:contours} for
a time interval spanning about 4 periods of the fundamental acoustic
mode. The black areas denote positive values for the convective flux
and therefore delimit the convective zone. An oscillatory behaviour is
clearly visible in each panel, with a period that looks similar to
the one of the fundamental acoustic mode, that is, almost 4 oscillation
cycles are depicted. This is consistent with the large peak shown around
$\omega_{00}$ in the power spectrum in Fig.~\ref{fig:fluxes}b.

The two TDC patterns also display a good agreement with the
nonlinear simulation regarding the overshooting phenomenon. We
indeed recover the same dark red structures below the convective
zone that correspond to the downdrafts entering in the radiative zone
(this penetration being associated with a negative convective flux).
Nevertheless, we note that the overshooting seems to be more vigorous
in Kuhfu\ss's model than in Stellingwerf's one as these dark
red structures fill in Fig.~\ref{fig:contours}c a larger surface in the bottom 
radiative zone than in Fig.~\ref{fig:contours}b.

\subsection{DNS vs TDC models: statistics of coefficients $\alpha$}
\label{stat}

\begin{figure}
 \centering
 \includegraphics[width=9cm]{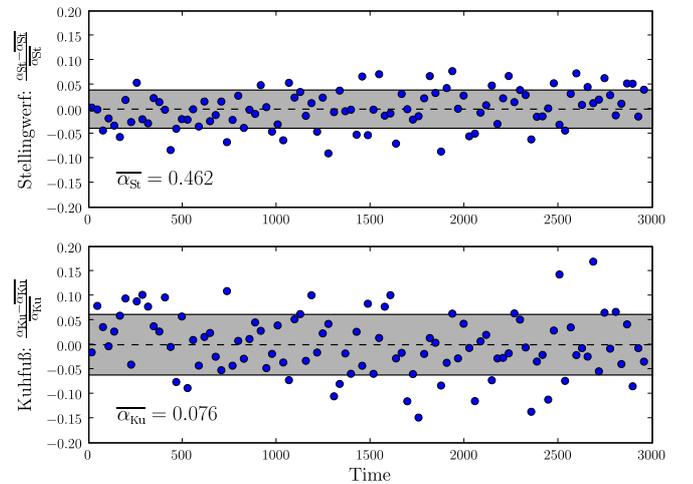}
 \caption{Time evolution around the mean value of coefficients
 $\alpst$ (\textit{upper panel}) and $\alpku$ (\textit{bottom panel}).
 Horizontal gray spans mark the limits of the associated relative
 standard deviation.}
\label{fig:alphas}
\end{figure}

A one-to-one comparison between the convective flux in the DNS and
its TDC predictions requires to find the optimal values of $\alpst$
and $\alpku$. This is done by performing several $\chi^2$-tests
at different snapshots in the simulation to track the variations of
coefficients $\alpha$. The resulting fluctuations of $\alpst$ and $\alpku$
around their mean values are shown in Fig.~\ref{fig:alphas}.

The dispersion of the Stellingwerf coefficient $\alpst$ (upper panel)
is weaker than the Kuhfu\ss~one (lower panel) as its values are almost within
a $5\%$ range around the mean
$\overline{\alpst} = 0.462$.
On the contrary, several outliers with values greater than
$10\%$ are found in the evolution of $\alpku$ which then appears more
chaotic around the mean value $\overline{\alpku}=0.076$. As a consequence,
the relative standard deviation (depicted in gray in Fig.~\ref{fig:alphas}) is
weaker in Stellingwerf's case than in Kuhfu\ss's one.

\subsection{DNS vs TDC models: convective fluxes with optimal $\alpha$'s}

\begin{figure}
 \centering
 \includegraphics[width=9cm]{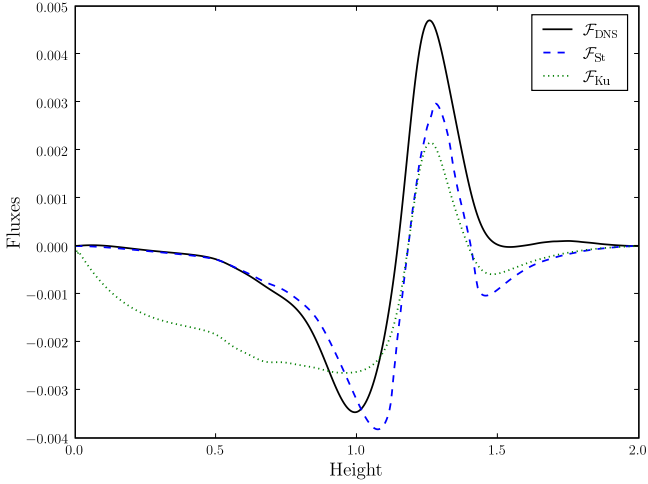}
 \caption{Mean convective flux in the DNS (solid black line), compared
 with the best TDC predictions based on the optimal $\alpha$ values that
 came out of the $\chi^2$-test in \S\ref{stat} (Stellingwerf's model in
 dashed blue line and Kuhfu\ss's one in dotted green line).}
\label{fig:fluxMLT}
\end{figure}

We recall that TDC models assume that the coefficients $\alpha$ entering
in Eq.~\ref{eq:MLT} are constant. By adjusting the TDC recipes with
the instantaneous convective flux throughout the nonlinear simulation,
the optimal value of these coefficients has been deduced. The final
test, given in Fig.~\ref{fig:fluxMLT}, then consists in the comparison
between the mean nonlinear convective flux taken over the entire
simulation and its best TDC approximations built from these optimal
$\alpha$ values.

This figure emphasises that Stellingwerf's formulation
gives a better agreement than Kuhfu\ss's one. Indeed, the
Kuhfu\ss~model overestimates the overshooting as the (negative) convective
flux remains non-negligible until the bottom of the radiative zone. On
the contrary, the Stellingwerf profile better accounts for the local 
penetration of convective plumes and shows the same exponential-like
decay of the negative convective flux when sinking in the
radiative zone \citep[e.g.][]{Dintrans2009}.
However, the two models are rather similar in the bulk of the convective
zone where convection is fully developed. One also notes that they both
predict a negative flux at the top of the convective zone, that is,
an upper overshooting of convection motions near the surface that is
not observed in the nonlinear simulation.

\section{Conclusion}

The main weakness of theories of time-dependent convection (TDC) lies in
the large number of free parameters. This is particularly awkward
when convection is strongly coupled with pulsations as, for example,
near the red border of the Cepheid instability strip where similar
results are obtained with different sets of parameters \citep{YKB98}.
Additional constraints must be found to reduce the intrinsic degeneracy
of models. But the modelling of the interplay between the turbulent
convection and oscillations is really a difficult task due to the strong
nonlinearities involved in this coupling.

One solution to tackle this problem and to bring new constraints
consists in performing fully nonlinear simulations, and this is
the path we have chosen in this work following our first study
in GD2011. Two widely used TDC theories, namely the \cite{St} and \cite{K} 
ones, are compared
with results coming from 2-D nonlinear simulations of compressible
convection in which strong acoustic oscillations are self-sustained
by the $\kappa$-mechanism. The heat transport is then modulated by the
fundamental acoustic mode such that this kind of simulation is
relevant to investigate the convection-pulsation coupling
(Figs.~\ref{fig:fluxes}-\ref{fig:contours}).

Focusing on the two TDC formulae for the convective flux, we compute the
evolution of free coefficients ``$\alpha$'' from a $\chi^2$-test applied
to the fully nonlinear results (Fig.~\ref{fig:alphas}). A large temporal
variability is found in both cases that weakens the robustness of
the TDC assumption $\alpha=\hbox{const}$. Moreover, the mean values
$\overline{\alpha}$ are not universal. By
applying the same method to other simulations performed in GD2011
(Table~\ref{tab:alphas}), we indeed do not recover the same
$\overline{\alpha}$ with a relative standard deviation of about
$12\%$ (Stellingwerf) and $18\%$ (Kuhfu\ss).

\begin{table}
 \centering
 \caption{Values of the optimal Stellingwerf and Kuhfu\ss~$\alpha$
 coefficients pulled out from the nonlinear simulations in GD2011.}
 \begin{tabular}{ccc}
 \toprule
 Simulation & $\overline{\alpst}$ & $\overline{\alpku}$ \\
 \midrule
 {\textbf{G8}} & 0.46 & 0.076\\
 G8H9 & 0.47 & 0.099 \\
 G8H8 & 0.33 & 0.113 \\
 G7   & 0.38 & 0.098 \\
 G6   & 0.38 & 0.102 \\
 G6F7 & 0.40 & 0.082 \\
 G6F5 & 0.38 & 0.067 \\
 \bottomrule
 \end{tabular}
 \label{tab:alphas}
\end{table}

Within these limits, Stellingwerf's formulation is found to give a better
agreement with the nonlinear simulations than Kuhfu\ss's one: ({\it i})
the final mean convective flux $\Fst$ is closer to its DNS counterpart,
with a much better estimation of the bottom overshooting; ({\it ii})
the temporal dispersion of the $\alpst$ coefficient is weaker, then its
enhanced stability (Fig.~\ref{fig:fluxMLT}). This result means that the
time-dependent convective flux better scales with a law $\Fc \propto \Et
\sqrt{\nabla -\gradad }$ than $\Fc \propto \sqrt{\Et}(\nabla -\gradad)$,
and the Stellingwerf formulation may probably be preferred in the 1-D
hydrocode used in, e.g., the topical Cepheids models. However, this study
emphasises that both formalisms lead to an artificial overshooting at
the top of the convection zone. We also note that this TDC test
involves the \emph{exact} value of the turbulent kinetic
energy $\Et$ provided by the nonlinear simulation, and not by the dedicated
1-D TDC equation which is inherently less accurate. As a consequence,
the obtained profiles are certainly the best we can expect from
these TDC recipes.

In this work, the temporal modulation of the convective flux is ensured
by the acoustic modes excited by $\kappa$-mechanism. An interesting
prospect could be the case of a modulation based on the internal
gravity waves excited by convection itself. Indeed, convection can
excite gravity waves in variable stars, either by the means of the
penetration of convective elements into stably stratified regions as
in solar-type stars \citep[e.g.][]{Din05}, or through the so-called
``convective blocking'' mechanism as in white dwarfs or Gamma Doradus
stars \citep[e.g.][]{Pes87}. In both cases, the convective flux is ipso
facto modulated by gravity waves and it may be interesting in that
respect to also check the accuracy of time-dependent convection models.

\begin{acknowledgements}
This work was granted access to the HPC resources of CALMIP under the
allocation 2010-P1021 (\url{http://www.calmip.cict.fr}).
\end{acknowledgements}

\nocite{*}
\bibliographystyle{aa}

\begin{thebibliography}{26}
\expandafter\ifx\csname natexlab\endcsname\relax\def\natexlab#1{#1}\fi

\bibitem[{{Baker} \& {Kippenhahn}(1965)}]{Baker65}
{Baker}, N. \& {Kippenhahn}, R. 1965, \apj, 142, 868

\bibitem[{{Baker}(1987)}]{Baker}
{Baker}, N.~H. 1987, in Physical Processes in Comets, Stars and Active
  Galaxies, ed. W.~{Hillebrandt}, E.~{Meyer-Hofmeister}, \& H.-C. {Thomas}
  (Berlin \& New York: Springer-Verlag), 105

\bibitem[{{B{\"o}hm-Vitense}(1958)}]{BV2}
{B{\"o}hm-Vitense}, E. 1958, Zeitschrift f\"ur Astrophysik, 46, 108

\bibitem[{{Bono} {et~al.}(1999){Bono}, {Marconi}, \& {Stellingwerf}}]{Bono}
{Bono}, G., {Marconi}, M., \& {Stellingwerf}, R.~F. 1999, \apjs, 122, 167

\bibitem[{{Bono} \& {Stellingwerf}(1994)}]{BS}
{Bono}, G. \& {Stellingwerf}, R.~F. 1994, \apjs, 93, 233

\bibitem[{{Brandenburg} \& {Dobler}(2002)}]{Pencil-Code}
{Brandenburg}, A. \& {Dobler}, W. 2002, CoPhC, 147, 471

\bibitem[{{Buchler}(2000)}]{Buchler-ASP}
{Buchler}, J.~R. 2000, ASPCS, 203, 343

\bibitem[{{Buchler}(2009)}]{Buchler09}
{Buchler}, J.~R. 2009, in AIPCS, ed. {J.~A.~Guzik \& P.~A.~Bradley}, Vol. 1170
  (Melville: American Institute of Physics), 51

\bibitem[{{Dintrans}(2009)}]{Dintrans2009}
{Dintrans}, B. 2009, CoAst, 158, 45

\bibitem[{{Dintrans} {et~al.}(2005){Dintrans}, {Brandenburg}, {Nordlund}, \&
  {Stein}}]{Din05}
{Dintrans}, B., {Brandenburg}, A., {Nordlund}, {\AA}., \& {Stein}, R.~F. 2005,
  \aap, 438, 365

\bibitem[{{Feuchtinger}(1999)}]{Feuch99}
{Feuchtinger}, M.~U. 1999, \aap, 136, 217

\bibitem[{{Gastine} \& {Dintrans}(2008{\natexlab{a}})}]{paperI}
{Gastine}, T. \& {Dintrans}, B. 2008{\natexlab{a}}, \aap, 484, 29

\bibitem[{{Gastine} \& {Dintrans}(2008{\natexlab{b}})}]{paperII}
{Gastine}, T. \& {Dintrans}, B. 2008{\natexlab{b}}, \aap, 490, 743

\bibitem[{{Gastine} \& {Dintrans}(2011)}]{paperIII}
{Gastine}, T. \& {Dintrans}, B. 2011, \aap, 528, A6, \textbf{[GD2011]}

\bibitem[{{Gehmeyr} \& {Winkler}(1992{\natexlab{a}})}]{GW}
{Gehmeyr}, M. \& {Winkler}, K.-H.~A. 1992{\natexlab{a}}, \aap, 253, 92

\bibitem[{{Gehmeyr} \& {Winkler}(1992{\natexlab{b}})}]{GW2}
{Gehmeyr}, M. \& {Winkler}, K.-H.~A. 1992{\natexlab{b}}, \aap, 253, 101

\bibitem[{{Gough}(1977)}]{Gough}
{Gough}, D.~O. 1977, ApJ, 214, 196

\bibitem[{{Koll{\'a}th} {et~al.}(2002){Koll{\'a}th}, {Buchler}, {Szab{\'o}}, \&
  {Csubry}}]{Koll02}
{Koll{\'a}th}, Z., {Buchler}, J.~R., {Szab{\'o}}, R., \& {Csubry}, Z. 2002,
  \aap, 385, 932

\bibitem[{{Kuhfu\ss}(1986)}]{K}
{Kuhfu\ss}, R. 1986, A\&A, 160, 116

\bibitem[{{Pesnell}(1987)}]{Pes87}
{Pesnell}, W.~D. 1987, \apj, 314, 598

\bibitem[{{Smolec} \& {Moskalik}(2010)}]{Smolec2010}
{Smolec}, R. \& {Moskalik}, P. 2010, \aap, 524, A40

\bibitem[{{Stellingwerf}(1982)}]{St}
{Stellingwerf}, R.~F. 1982, ApJ, 262, 330

\bibitem[{{Unno}(1967)}]{Unno}
{Unno}, W. 1967, PASJ, 19, 140

\bibitem[{{Vitense}(1953)}]{BV1}
{Vitense}, E. 1953, Zeitschrift f\"ur Astrophysik, 32, 135

\bibitem[{{Wuchterl} \& {Feuchtinger}(1998)}]{WF}
{Wuchterl}, G. \& {Feuchtinger}, M.~U. 1998, A\&A, 340, 419

\bibitem[{{Yecko} {et~al.}(1998){Yecko}, {Koll{\'a}th}, \& {Buchler}}]{YKB98}
{Yecko}, P.~A., {Koll{\'a}th}, Z., \& {Buchler}, J.~R. 1998, \aap, 336, 553

\end{thebibliography}

\Online
\twocolumn[\vspace*{4cm}\noindent{\sffamily\bfseries\Huge\centerline{Online
Material}}]
\vfill
\clearpage

\begin{appendix}

\section{Setup of the simulations}

\subsection{The equilibrium model}

\begin{table*}[h!]
 \centering
 \caption{Dimensionless parameters of the numerical simulations.}
\begin{tabular}{cccccccccc}
 \toprule
 DNS & Gravity $\vec{g}$ & Flux $\Fbot$ & Conductivity $\Kmax$ & $\Tbump$ &
Width $e$ & Slope $\sigma$ &Viscosity $\nu$ & Frequency $\omega_{00}$ &
Rayleigh\\
 \midrule
 {\textbf{G8}} & 8 & $4.5\times 10^{-2}$ & $10^{-2}$ & 6
&1 & 1&$2.5\times 10^{-4}$ & $3.85$ & $1 0^5$\\
G8H9 & 8 & $4.5\times 10^{-2}$ & $9\times 10^{-3}$ & 7 & 1 & 1 & $5\times
10^{-4}$ & $3.83$ & $10^{5} $\\
G8H8 & 8 & $4.5\times 10^{-2}$ & $8\times
10^{-3}$& 7.5 & 1 & 1.1 & $5\times 10^{-4}$ & $3.80$ & $2\times 10^{5}$ \\
G7  & 7 & $4\times 10^{-2}$ & $10^{-2}$ & 5.5 & 1.5 & 0.8 & $2.5\times 10^{-4}$
& $3.62$ & $8\times 10^{4} $ \\
G6   & 6 & $4\times 10^{-2}$ & $10^{-2}$ & 6 & 1.5 & 0.8 & $5\times 10^{-4}$ &
$3.35$ & $8\times 10^{4} $ \\
G6F7 & 6 & $3.7\times 10^{-2}$ & $10^{-2}$ & 5.7 & 1.5 & 0.8 & $3.5\times
10^{-4}$ & $3.36$ & $9\times 10^{4} $ \\
G6F5 & 6 & $3.5\times 10^{-2}$ & $10^{-2}$ & 5.5 & 1.5 & 0.8 & $3\times 10^{-4}$
& $3.36$ & $9\times 10^{4} $ \\ 
\bottomrule
 \end{tabular}
\tablefoot{%
The bold-typed one emphasises the DNS mainly discussed in this study.
For all these simulations, we assume $\Ttop=2$, $\rhotop=10^{-2}$, $\Amp=0.7$,
and $L_x/L_z=4$.}
 \label{tab:DNS}
\end{table*}

As in GD2011, our system represents a zoom around an ionisation region.
As we are computing \emph{local} simulations, the vertical gravity
$\vec{g}=-g\vec{e_z}$ and the kinematic viscosity $\nu$ are assumed to be
constant. Following our purely radiative model of the $\kappa$-mechanism
\citep{paperI,paperII}, the ionisation region is represented by a
temperature-dependent radiative conductivity
profile that mimics an opacity bump:
  
\begin{equation}
 K(T)=\Kmax\left[1+\Amp\dfrac{-\pi/2+\arctan(\sigma
T^+T^-)}{\pi/2+\arctan(\sigma e^2)}\right],
\label{eq:conductivity-profile1}
\end{equation}
with

\begin{equation}
 \Amp=\dfrac{\Kmax-\Kmin}{\Kmax},\quad T^{\pm}=T-\Tbump \pm e,
\label{eq:conductivity-profile2}
\end{equation}
where $\Tbump$ is the position of the hollow in temperature and
$\sigma$, $e$, and $\Amp$ denote its slope, width, and relative amplitude,
respectively. We assume both radiative and hydrostatic equilibria; that is,

\begin{equation}
 \left\lbrace
 \begin{aligned}
  \dfrac{dp_0}{dz} & = -\rho_0 g, \\
  \dfrac{dT_0}{dz} & = -\dfrac{\Fbot}{K_0(T_0)}, \\
 \end{aligned}
 \right.
 \label{eq:equil}
\end{equation}
where $\Fbot$ is the imposed bottom flux. Following GD2011, we chose
$L_z$ as the length scale, i.e. $[x]=L_z$, top density $\rhotop$ and 
top temperature $\Ttop$ as density and temperature scales,
respectively. The velocity scale is then $\sqrt{c_p \Ttop}$, while time is
given in units of $[t] = L_z/ \sqrt{c_p \Ttop}$.

Table \ref{tab:DNS} then summarises in these dimensionless units the parameters
of the numerical simulations presented in this study. The penultimate column of
this table contains the value of the frequency $\omega_{00}$ of the fundamental
unstable radial mode excited by the $\kappa$-mechanism, which lies between 3 and
4 for every DNS. The last column gives the value of the Rayleigh number, which
quantifies the strength of the convective motions. It is given by

\begin{equation}
 \text{Ra} = \dfrac{g L_{\text{conv}}^4}{\nu \chi
c_p}\left|\dfrac{ds}{dz}\right|,
 \label{eq:rayleigh}
\end{equation}
where $L_{\text{conv}}$ is the width of the convective zone,
$\chi=K_0/\rho_0 c_p$ the radiative diffusivity, and $s$ the entropy.

\subsection{The nonlinear equations}

With the parallel alternate direction implicit solver for the radiative
diffusion implemented in the pencil code (see GD2011), we advance the following
hydrodynamic equations in time:
\begin{equation}
\left\lbrace
\begin{aligned}
\disp \frac{D \ln \rho}{Dt} = & -\Div \vec{u}, \\ \\
\disp \frac{D \vec{u}}{Dt} = & - \dfrac{1}{\rho}\na p
+ \vec{g} + 2\nu \lp \na \cdot \vec{\sf S} +\vec{\sf S}\cdot \na\ln\rho\rp, \\
\\
\disp \frac{DT}{Dt} = & \frac{1}{\rho c_v} \Div K(T)\na T -(\gamma
-1)T\Div\vec{u
} + 2\rho\nu \vec{\sf S}^2,
\end{aligned}
\right.
\label{eq:hydro}
\end{equation}
where $\rho,\ \vec{u}$, and $T$ denote density, velocity, and
temperature, respectively, while $K(T)$ is given by
Eq.~\ref{eq:conductivity-profile1}. The operator $D/Dt = \partial / \partial t +
\vec{u} \cdot \na$ is the usual total derivative, while $\vec{\sf S}$ is the
(traceless) rate-of-strain tensor given by

\begin{equation}
{\sf S}_{ij} = \frac{1}{2}\left(\frac{\partial u_i}{\partial x_j} +
\frac{\partial u_j}{\partial x_i}-\frac{2}{3} \delta_{ij} \Div \vec{u}\right).
\end{equation}
Finally, we impose that all fields are periodic in the horizontal
direction, while stress-free boundary conditions (i.e. $u_z=0$ and
$du_x/dz=0$) are assumed for the velocity in the vertical one. Concerning
the temperature, a perfect conductor at the bottom (i.e. flux imposed)
and a perfect insulator at the top (i.e. temperature imposed) are applied.

In order to ensure that both the nonlinear saturation and thermal
relaxation are achieved, the simulations were computed over very long times,
typically $t\gtrsim 3000$. As the eigenfrequency of the unstable acoustic mode
$\omega_{00} \in [3-4]$ (see Table~\ref{tab:DNS}), this corresponds
approximately to 1800 periods of oscillation.

\end{appendix}
 
\end{document}